\documentclass[twocolumn,preprintnumbers,showpacs,amsmath,amssymb,prb]{revtex4}

\usepackage{graphicx}
\usepackage{dcolumn}
\usepackage{bm}
\usepackage{amssymb}
\begin{document}

\title{Magnetoelectronic properties of graphene dressed by a high-frequency field}
\author{O. V. Kibis$^{1,2}$}\email{Oleg.Kibis(c)nstu.ru}
\author{S. Morina$^{2,3}$}
\author{K. Dini$^2$}
\author{I. A. Shelykh$^{2,3,4}$}
\affiliation{${^1}$Department of Applied and Theoretical Physics,
Novosibirsk State Technical University, Karl Marx Avenue 20,
Novosibirsk 630073, Russia} \affiliation{$^2$Science Institute,
University of Iceland, Dunhagi 3, IS-107, Reykjavik, Iceland}
\affiliation{$^3$Division of Physics and Applied Physics, Nanyang
Technological University 637371, Singapore} \affiliation{$^4$ITMO
University, Saint Petersburg 197101, Russia}

\begin{abstract}
Solving the Schr\"odinger problem for electrons in graphene
subjected to both a stationary magnetic field and a strong
high-frequency electromagnetic wave (dressing field), we found
that the dressing field drastically changes the structure of
Landau levels in graphene. As a consequence, the magnetoelectronic
properties of graphene are very sensitive to the dressing field.
Particularly, it is demonstrated theoretically that the dressing
field strongly changes the optical spectra and the Shubnikov-de
Haas oscillations. As a result, the developed theory opens a way
for controlling magnetoelectronic properties of graphene with
light.
\end{abstract}

\maketitle

\section{Introduction} Since the discovery of graphene
\cite{Novoselov_04}, its unique electronic properties have aroused
enormous interest in the scientific community
\cite{CastroNeto_2009,DasSarma_2011,Goerbig}. Particularly, the
magnetoelectronic properties of graphene --- effects caused by the
influence of a stationary magnetic field on the electron energy
spectrum \cite{Yang,Jiang,Zhang,Sadowski}, optical characteristics
\cite{Yao,Booshehri,Crassee,Crasseebis,Grujic,Shimano,Yao_13} and
electronic transport \cite{Ando_98,
Guinea,Das_Sarma,Tan,Qiao,Waldmann,Waldmannbis,Dean}
--- are in the focus of attention. Since a magnetic field effectively
controls electronic properties of graphene, studies on the subject
are important from viewpoint of both fundamental physics and
graphene-based electronics. Besides a stationary magnetic field,
an effective tool to manipulate electronic properties is a strong
high-frequency electromagnetic field. Since the system ``electron
+ strong electromagnetic field'' should be considered as a whole,
the bound electron-field object --- ``electron dressed by
electromagnetic field'' (dressed electron) --- became a commonly
used model in modern physics
\cite{Cohen-Tannoudji_b98,Scully_b01}. The physical properties of
dressed electrons have been studied in both atomic systems
\cite{Autler_55,Cohen-Tannoudji_b98,Scully_b01} and various
condensed-matter structures, including bulk semiconductors
\cite{Elesin_69,Vu_04,Vu_05}, quantum wells
\cite{Mysyrovich_86,Wagner_10,Kibis_12,Teich_13,Kibis_14,Morina_15,Pervishko_15},
quantum rings \cite{Kibis_11,Kibis_14_1,Joibari_14,Kyriienko_15},
etc. In graphene, a dressing field can strongly modify both the
electron energy spectra and electronic transport
\cite{Lopez_08,Oka_09,Kitagawa_11,Kibis_10,Kibis_11_1,Usaj_14,Glazov_14,Lopez_15}.
Particularly, magneto-like electronic effects (so-called
photovoltaic Hall effect, etc) can be induced by a dressing field
in the absence of a stationary magnetic field
\cite{Oka_09,Kitagawa_11}. Therefore, one can expect that the
magnetoelectronic properties of graphene are strongly affected by
a dressing field as well. However, a consistent theory describing
magnetoelectronic properties of dressed graphene was not
elaborated up to now. The present paper is aimed to fill partially
this gap.

\section{Model} To describe the magnetoelectronic
properties of dressed graphene, we have to solve the Schr\"odinger
problem for electrons in a graphene layer exposed to both an
electromagnetic wave (dressing field) and a stationary magnetic
field (see Fig.~1).
\begin{figure}
\includegraphics[scale=0.7]{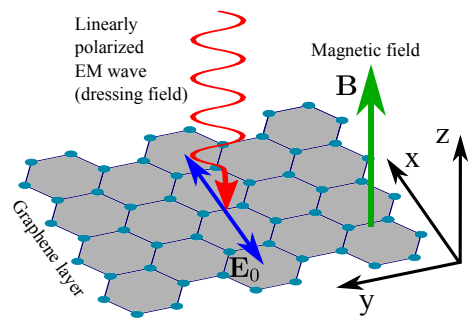}
\caption{\label{fig:figure1} (Color online) Sketch of the system
under consideration: A graphene layer subjected to both a linearly
polarized electromagnetic wave (EM) with the electric field
amplitude $E_0$ and a stationary magnetic field, $\mathbf{B}$,
directed perpendicularly to the layer.}
\end{figure}
Generally, electronic states near the $K$ and $K^\prime$ points of
the Brillouin zone of graphene (the Dirac points) can be described
by eight-component wave functions written in a basis corresponding
to two crystal sublattices of graphene, two electron valleys, and
two orientations of electron spin \cite{CastroNeto_2009}. In the
following, the intervalley mixing of electron states and spin
effects will be beyond consideration. Therefore, the number of
necessary wave-function components can be reduced to two. Within
this conventional approximation, the Hamiltonian of electrons near
the $K$ point of the Brillouin zone has the form
\begin{equation}\label{H0}
\hat{{\cal H}}=
v{\bm{\sigma}}\cdot\left[\hat{\mathbf{p}}+\frac{e}{c}
\left({\mathbf{A}}_0+\mathbf{A}_B\right)\right],
\end{equation}
where $v$ is the electron velocity at the Dirac point,
$\hat{\mathbf{p}}=(\hat{p}_x,\hat{p}_y)$ is the operator of
electron momentum in the graphene layer, $e$ is the modulus of
electron charge, ${\mathbf{A}}_0=([c{E}_0/\omega]\cos\omega
t,0,0)$ is the vector potential of the linearly polarized
electromagnetic wave propagating perpendicularly to the graphene
plane, $E_0$ is the amplitude of electric field of the wave,
$\omega$ is the wave frequency, $\mathbf{A}_B=(-By,0,0)$ is the
vector potential of the stationary magnetic field,
$\mathbf{B}=(0,0,B)$, which is assumed to be directed
perpendicularly to the graphene layer, and
${\bm{\sigma}}=(\sigma_x,\sigma_y,\sigma_z)$ is the vector of
Pauli matrices written in the basis of two orthogonal electron
states arisen from the two crystal sublattices of
graphene~\cite{CastroNeto_2009}. Formally, these two basis states,
$|+\rangle$ and $|-\rangle$, correspond to the two opposite
orientations of the pseudospin along the $z$-axis,
$\sigma_z|\pm\rangle=\pm|\pm\rangle$.

Let us introduce the new two orthonormal states,
\begin{equation}\label{psi0}
\psi_0^\pm=\frac{|+\rangle\pm|-\rangle}{\sqrt{2}}e^{\mp
i(\alpha/2)\sin\omega t},
\end{equation}
where $\alpha={2veE_0}/{\hbar\omega^2}$ is the dimensionless
parameter describing the interaction between an electron in
graphene and a dressing field. Since Eq.~(\ref{psi0}) defines the
complete system of basis states of graphene at any time $t$, one
can seek solutions of the Schr\"odinger problem as
\begin{equation}\label{psi}
\psi(\mathbf{r},t)=a^+(\mathbf{r},t)\psi_0^++a^-(\mathbf{r},t)\psi_0^-,
\end{equation}
where $\mathbf{r}=(x,y)$ is the electron radius-vector in the
graphene plane. Substituting the wave function (\ref{psi}) into
the Schr\"odinger equation with the Hamiltonian (\ref{H0}),
$i\hbar{\partial\psi(\mathbf{r},t)}/{\partial
t}=\hat{\cal{H}}\psi(\mathbf{r},t)$, we arrive at the two
differential equations,
\begin{eqnarray}\label{a}
i\hbar\dot{a}^\pm(\mathbf{r},t)&=&\pm
v\left[\hat{p}_x-\frac{eBy}{c}\right]
{a}^\pm(\mathbf{r},t)\nonumber\\
&\pm& ive^{\pm i\alpha\sin\omega t}\hat{p}_y{a}^\mp(\mathbf{r},t),
\end{eqnarray}
which describe the quantum dynamics of dressed electrons in the
graphene layer. Applying the conventional Floquet theory of
periodically driven quantum
systems~\cite{Zeldovich_67,Grifoni_98,Platero_04} to the wave
function (\ref{psi}), we can rewrite it as
$\psi(\mathbf{r},t)=e^{-i\tilde{\varepsilon}
t/\hbar}\phi(\mathbf{r},t)$, where the function
$\phi(\mathbf{r},t)$ periodically depends on time,
$\phi(\mathbf{r},t)=\phi(\mathbf{r},t+2\pi/\omega)$, and
$\tilde{\varepsilon}$ is the quasi-energy of an electron. Since
the quasi-energy (the energy of dressed electron) is the physical
quantity which plays the same role in periodically driven quantum
systems as an usual energy in stationary ones, the present
analysis of the Schr\"odinger problem is aimed to find the energy
spectrum of dressed electron, $\tilde{\varepsilon}$. Taking into
account the periodicity of the function $\phi(\mathbf{r},t)$, one
can seek the coefficients ${a}^\pm(\mathbf{r},t)$ in Eq.~(\ref{a})
as a Fourier expansion,
\begin{equation}\label{aF}
{a}^\pm(\mathbf{r},t)=e^{-i\tilde{\varepsilon}
t/\hbar}\sum_{n=-\infty}^{\infty}a_{n}^\pm(\mathbf{r})e^{in\omega
t}.
\end{equation}
Substituting the expansion (\ref{aF}) into the expression
(\ref{a}) and applying the Jacoby-Anger expansion, $
e^{iz\sin\theta}=\sum_{n=-\infty}^{\infty}J_n(z)e^{in\theta}, $ to
transform the exponent in the right side, one can rewrite
Eq.~(\ref{a}) as
\begin{eqnarray}\label{a1}
(\tilde{\varepsilon}-n\hbar\omega){a}^\pm_n(\mathbf{r})&=&\pm v\left[\hat{p}_x-\frac{eBy}{c}\right]{a}^\pm_n(\mathbf{r})\nonumber\\
&\pm& iv\sum_{n^\prime=-\infty}^\infty
J_{n-n^\prime}\left(\pm\alpha\right)
\hat{p}_y{a}^\mp_{n^\prime}(\mathbf{r}).
\end{eqnarray}
It should be noted that Eq.~(\ref{a1}) still describes exactly the
initial Schr\"odinger problem. Next we will make some
approximations.

Let us assume that the wave frequency, $\omega$, is far from
resonant electron frequencies corresponding to electron
transitions between the different Landau levels in graphene, and,
therefore, the inter-level absorption of the wave by electrons is
absent. Thus, the considered electron system is conservative.
Next, we have to take into account that the expansion coefficients
in Eq.~(\ref{aF}), ${a}^\pm_{n}(\mathbf{r})$, are the quantum
amplitudes of the absorption (emission) of $n$ photons by an
electron. Since the considered nonresonant field can be neither
absorbed nor emitted by an electron, the amplitudes are very
small, $\left|{a}^\pm_{n\neq0}(\mathbf{r})\right|\ll1$. Assuming
the zero-order Bessel function, $J_0(\alpha)$, to be far from
zero, the aforesaid leads to the estimation
\begin{equation}\label{cond}
\left|\frac{J_n(\alpha)\hat{p}_y{a}^\pm_{n}(\mathbf{r})}{J_0(\alpha)\hat{p}_y{a}^\pm_{0}(\mathbf{r})}\right|\ll1,
\end{equation}
where $n=\pm1,\pm2,...$ It follows from the inequality
(\ref{cond}) that the main contribution to the sum in
Eq.~(\ref{a1}) arises from terms with $n^\prime=0$, which describe
the elastic interaction between an electron and the dressing
field. Therefore, small terms with ${a}^\pm_{n\neq0}(\mathbf{r})$
in Eq.~(\ref{a1}) can be omitted. It should be noted that such a
neglect of high-frequency nonresonant terms in Eq.~(\ref{a1}) is
physically identical to the rotating wave approximation (RWA)
which is conventionally used to describe various quantum systems
under periodical pumping (see, e.g.,
Refs.~\cite{Cohen-Tannoudji_b98,Scully_b01}). Within this
approach, Eq.~(\ref{a1}) turns into the equation
\begin{equation}\label{a2}
\tilde{\varepsilon}{a}^\pm_0(\mathbf{r})=\pm
v\left[\hat{p}_x-\frac{eBy}{c}\right]{a}^\pm_0(\mathbf{r}) \pm
ivJ_{0}\left(\alpha\right) \hat{p}_y{a}^\mp_{0}(\mathbf{r}).
\end{equation}
Formally, Eq.~(\ref{a2}) can be treated as a stationary
Schr\"odinger equation, $\hat{\cal H}_0
a_0(\mathbf{r})=\tilde{\varepsilon}a_0(\mathbf{r})$, with the
effective Hamiltonian
\begin{equation}\label{H1}
\hat{\cal H}_0=\sigma_z
v\left[\hat{p}_x-\frac{eBy}{c}\right]-\sigma_y
vJ_{0}\left(\alpha\right)\hat{p}_y,
\end{equation}
where $a_0(\mathbf{r})$ is the pseudospinor with the two
components, $a^\pm_0(\mathbf{r})$. Applying the unitary
transformation, $\hat{U}=(\sigma_z+\sigma_x)/\sqrt{2}$, to the
Hamiltonian (\ref{H1}), we arrive at the transformed Hamiltonian,
$\hat{\cal H}_0^\prime=\hat{U}^\dagger\hat{\cal H}_0\hat{U}$,
which has the well-behaved form
\begin{equation}\label{H2}
\hat{\cal H}_0^\prime=\sigma_x
\tilde{v}_x\left[\hat{p}_x-\frac{eBy}{c}\right]+\sigma_y
\tilde{v}_y\hat{p}_y,
\end{equation}
where the quantities $\tilde{v}_x=v$ and
$\tilde{v}_y=vJ_0(\alpha)$ should be treated as components of the
velocity of dressed electron along the $x,y$ axes. It should be
noted that the electron velocity along the polarization vector of
the dressing field, $\tilde{v}_x=v$, is not changed by the the
dressing field, whereas the electron velocity in the perpendicular
direction, $\tilde{v}_y=vJ_0(\alpha)$, drastically depends on the
field because of the Bessel-function factor.

If the magnetic field is absent, $B=0$, the Hamiltonian (\ref{H2})
can be diagonalized trivially and results in the anisotropic
energy spectrum of dressed electrons \cite{Kristinsson_16},
\begin{equation}\label{E0}
\tilde{\varepsilon}=\pm\hbar v\sqrt{k_x^2+k_y^2J_0^2(\alpha)},
\end{equation}
where $\mathbf{k}=(k_x,k_y)$ is the electron wave vector in the
graphene plane. If a graphene layer is exposed to the magnetic
field, $B\neq0$, the Hamiltonian (\ref{H2}) is mathematically
identical to the known Hamiltonian of ``bare'' graphene subjected
to the same magnetic field, where the velocity of ``bare''
electron, $\mathbf{v}=(v,v)$, should be replaced with the velocity
of dressed electron,
$\tilde{\mathbf{v}}=(\tilde{v}_x,\tilde{v}_y)$. As a consequence,
the electron eigenenergies and eigenfunctions corresponding to the
Landau levels in dressed graphene can be easily obtained from
well-known those for ``bare'' graphene \cite{CastroNeto_2009} with
the formal replacement, $v\rightarrow v\sqrt{|J_0(\alpha)|}$.
Particularly, the energies of the Landau levels in dressed
graphene read as
\begin{equation}\label{En}
\tilde{\varepsilon}_n=\mathrm{sgn}(n)\hbar\omega_B\sqrt{|n|}
\sqrt{|J_0(\alpha)|},
\end{equation}
where $\omega_B=\sqrt{2}v/l_B$ is the cyclotron frequency of
graphene, $l_B=\sqrt{\hbar c/eB}$ is the magnetic length,
$\mathrm{sgn}(n)$ is the signum function, and $n=0,\pm1,\pm2...$
is the number of Landau level in the conductivity band ($n>0$) and
the valence band ($n<0$). As expected, the energies (\ref{En})
exactly coincides with those in ``bare'' graphene
\cite{CastroNeto_2009} if the dressing field is absent
($\alpha=0$). To avoid misunderstandings, one should keep in mind
that the present theory is elaborated under condition
(\ref{cond}). Therefore, Eqs.~(\ref{a2})--(\ref{En}) are relevant
if the Bessel function, $J_0(\alpha)$, is far from zero.

According to Eq.~(\ref{En}), the dressing field changes the
distance between the Landau levels. Physically, this effect
originates from the linear electron dispersion in graphene.
Indeed, in conducting systems with the parabolic dispersion of
electrons the Landau levels are stable against a dressing field:
The dressing field shifts the Landau levels uniformly but does not
change the distance between them \cite{Inoshita_00}. Therefore,
the magnetoelectronic properties of graphene will be very
sensitive to the dressing field in contrast to the case of usual
conducting systems with the parabolic dispersion of electrons.

\section{Optical and transport effects} The field-induced
modification of the Landau levels (\ref{En}) will manifest itself
in various magneto-optical and magneto-transport phenomena. For
definiteness, we will focus the attention on the optical
absorption and longitudinal conductivity of dressed graphene.

Let dressed graphene be subjected to a weak linearly polarized
electromagnetic wave with the frequency $\Omega$ (probing field).
The probing field can induce electron transitions between the
dressed Landau levels (\ref{En}) which are accompanied by
absorption of the field. Conventionally, this optical effect can
be described by the absorption coefficient, $\beta$. Combining the
theory elaborated above and the known theory of magneto-optical
absorption in ``bare'' graphene (see, e.g., Ref.~\cite{Yao_13}),
we arrive at the expression for the absorption coefficient in
dressed graphene,
\begin{equation}\label{beta}
\beta_{i}=\sum_{n,m}\frac{2e^2\tilde{v}_{i}^2
\gamma({\rho_m-\rho_n})(1+\delta_{n,0}+\delta_{m,0})\delta_{|n|-|m|,\pm1}}
{l_B^2\hbar
c\tilde{\omega}_{nm}[(\tilde{\omega}_{nm}-\Omega)^2+\gamma^2]},
\end{equation}
where $\rho_{n}$ is the equilibrium filling factor for the Landau
levels (\ref{En}),
$\tilde{\omega}_{nm}=(\tilde{\varepsilon}_n-\tilde{\varepsilon}_m)/\hbar$
are the resonance frequencies of dressed graphene corresponding to
electron transitions between different Landau levels (\ref{En}),
$\gamma=\Gamma/\hbar$ is the decay rate at Landau levels which is
assumed to be independent on the irradiation, $\Gamma$ is the
scatterer-induced broadening of Landau levels, $\delta_{nm}$ is
the Kronecker symbol, and the index $i=x,y$ corresponds to the two
polarizations of the probing field along the $x,y$ axes,
respectively. The absorption coefficient (\ref{beta}) is plotted
in Fig.~2 for various intensities of the dressing field,
$I_0=cE_0^2/8\pi$.
\begin{figure}
\includegraphics[scale=0.6]{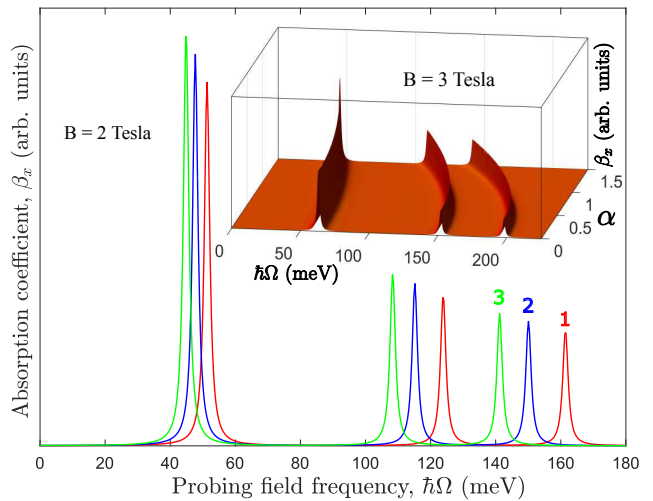}
\caption{\label{fig:figure2} (Color online) The absorption
coefficient of intrinsic graphene, $\beta_x$, at the temperature
$T=0$ as a function of the probing field frequency, $\Omega$, for
the decay rate $\Gamma=1$~meV. The photon energy of dressing field
is $\hbar\omega=3$~meV and the different curves correspond to the
different field intensities: (1) $I_0=0$; (2) $I_0=3.51$~W/cm$^2$;
(3) $I_0=6.24$~W/cm$^2$. The insert shows the dependence of the
absorption coefficient, $\beta_x$, on the photon energy,
$\hbar\Omega$, and the electron-field interaction parameter,
$\alpha=2evE_0/\hbar\omega^2$.}
\end{figure}

In order to analyze the longitudinal conductivity of dressed
graphene in the presence of a magnetic field, let us use the
conventional formalism based on the Kubo formula \cite{Ando_98}.
Within this approach, the diagonal components of the conductivity
tensor read as
\begin{equation}\label{Kubo1}
{\sigma}_{ii}=\int\mathrm{d}\varepsilon\left[-\frac{\partial
f(\varepsilon)}{\partial\varepsilon}\right]{\sigma}_{ii}(\varepsilon),
\end{equation}
with
\begin{equation}\label{Kubo}
{{\sigma}_{ii}(\varepsilon)}=\frac{e^2\hbar}{\pi
S}\mathrm{Tr}\langle
\hat{v}_i\,\mathrm{Im}\,G(\varepsilon+i0)\hat{v}_i\,\mathrm{Im}\,G(\varepsilon+i0)\rangle,
\end{equation}
where $\varepsilon$ is the electron energy, $f(\varepsilon)$ is
the Fermi distribution function, $S$ is the area of the graphene
layer, $\hat{v}_{x,y}$ is the operator of electron velocity along
the $x,y$ axes, $G(\varepsilon)$ is the Green's function of the
effective Hamiltonian (\ref{H2}), and the broken brackets in
Eq.~(\ref{Kubo}) correspond to averaging over all possible
configuration of random distributions of scatterers. Applying the
conventional Green's function technique within the self-consistent
Born approximation, Shon and Ando calculated the conductivity of
graphene for the cases of short-gange scatterers and long-range
ones in the most general form \cite{Ando_98}. Particularly, they
demonstrated that the self-energy is the same for the both kinds
of scatterers. As to the vertex corrections, they vanish in the
case of short-range scatterers but should be taken into account
for long-range ones. In order to incorporate a dressing field into
this known approach, we have to just replace the electron velocity
in ``bare'' graphene, $\mathbf{v}$, with the velocity renormalized
by the dressing field, $\tilde{\mathbf{v}}$, in the expression
(\ref{Kubo}). It should be noted that the conductivity
(\ref{Kubo}) can be calculated analytically in the important case
of weak scattering, when the scatterer-induced broadeining of
Landau levels, $\Gamma$, is much less than the energy interval
between nearest Landau levels. Taking into account the
contribution of only Landau level at the Fermi energy and assuming
the scattering processes in graphene to be caused by a long-range
``white noise'' disorder, we can write the conductivity
(\ref{Kubo}) as
\begin{equation}\label{sigma}
\sigma_{xx}(\varepsilon)=\sigma_0|n|\left[1-\frac{\pi|\varepsilon|(\varepsilon-\tilde{\varepsilon}_n)^2}{2\Gamma(\hbar\omega_B)^2}\right],
\end{equation}
where $\sigma_0=e^2/\pi^2\hbar$ is the conductivity of intrinsic
graphene, and $n=\pm1,\pm2,\pm3...$ is the number of Landau level
at the Fermi energy. As to the case of short-range disorder, it is
described by the same expression (\ref{sigma}) which should be
just mupltiplied by the factor 2. As expected, the conductivity
(\ref{sigma}) exactly coincides with the known expression
\cite{Ando_98} if a dressing field is absent ($E_0=0$). The
Shubnikov-de Haas oscillations of conductivity calculated within
the approach \cite{Ando_98} for a graphene layer with short-range
disorder are presented in Fig.~3 for various parameters of the
dressing field.
\begin{figure}
\includegraphics[scale=0.33]{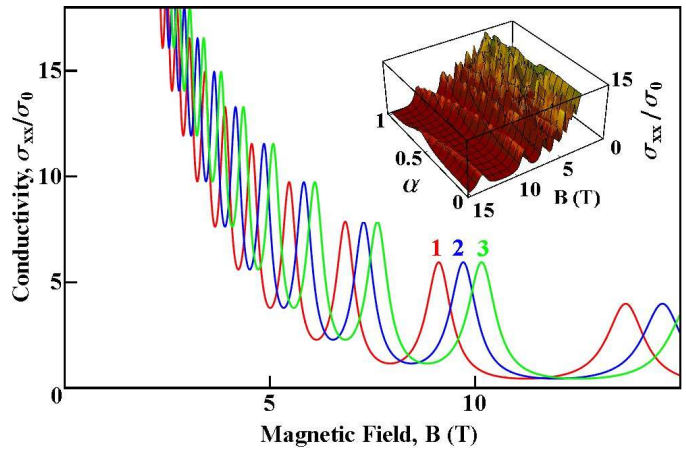}
\caption{\label{fig:figure3} (Color online) The conductivity of a
graphene layer, $\sigma_{xx}$, at the temperature $T=0$ as a
function of the magnetic field, $B$, for the Landau level
broadening $\Gamma=1$~meV and the Fermi energy
$\varepsilon_F=10$~meV. The photon energy of dressing field is
$\hbar\omega=3$~meV and the different curves correspond to the
different field intensities: (1) $I_0=0$; (2) $I_0=1.56$~W/cm$^2$;
(3) $I_0=2.63$~W/cm$^2$. The insert shows the dependence of the
conductivity, $\sigma_{xx}$, on the magnetic field, $B$, and the
electron-field interaction parameter,
$\alpha=2evE_0/\hbar\omega^2$.}
\end{figure}

\section{Discussion and conclusions} Deriving the effective
Hamiltonian (\ref{H2}), we assumed that the electromagnetic wave
is nonresonant. This allowed us to neglect the inter-level
absorbtion of the wave by electrons. However, we took also into
account the scatterring-induced broadening of Landau levels,
$\Delta=\hbar/\tau$, where $\tau$ is the electron scattering time.
For the self-consistency of the developed theory, we need to
exclude the scattering-induced intra-level absorption of the wave
by electrons within a broadened Landau level. Therefore, we have
to assume the wave frequency, $\omega$, to be high enough in order
to satisfy the inequality $\omega\tau\gg1$.  It is well-known that
the nonresonant (collisional) absorption of wave energy by
conduction electrons is negligibly small under this condition.
Therefore, an electromagnetic wave which is both high-frequency
and nonresonant can be treated as a purely dressing
(nonabsorbable) field (see, e.g.,
Refs.~\cite{Kibis_14,Morina_15,Pervishko_15}). Such a purely
dressing field should be used in experiments.

Physically, the absorption peaks in Fig.~2 correspond to the
resonant electron transitions between dressed Landau levels, which
are induced by a probing field. As to the Shubnikov-de Haas
oscillations of the conductivity plotted in Fig.~3, their maxima
correspond to the crossing of the Landau levels and the Fermi
level. Since the distance between Landau levels (\ref{En}) depends
on a dressing field, the field-induced shifting of the maxima of
the curves plotted in Figs.~2--3 appears. Besides the shifting,
the dressing field leads to the anisotropy of magnetoelectronic
properties caused by the nonequivalence of the electron velocities
along the $x,y$ axes, $\tilde{v}_x\neq\tilde{v}_y$. Namely, it
follows from Eqs.~(\ref{beta}) and (\ref{Kubo}) that
$\beta_y/\beta_x=\sigma_{yy}/\sigma_{xx}=|J_0(\alpha)|$. It should
be noted that the discussed field-induced effects are most
pronounced if the parameter of electron-field interaction,
$\alpha={2veE_0}/{\hbar\omega^2}$ is large enough. Due to the
giant electron velocity in graphene, $v\approx1\cdot10^6$~m/s,
this condition can be satisfied for relatively weak intensities of
the dressing field (see Figs.~2--3). As a consequence, the optical
and transport measurements are appropriate to detect the
field-induced modification of electron energy spectrum (\ref{En})
for experimentally reasonable parameters of the dressing field.

Physically, the considered features of the magnetoelectronic
properties originate from the anisotropy of the energy spectrum of
dressed electrons (\ref{E0}) along the $x,y$ axes. In turn, the
anisotropy is caused by linear polarization of a dressing field.
In the case of circularly polarized dressing field with the same
electric field amplitude $E_0$ and frequency $\omega$, the
anisotropic spectrum (\ref{E0}) turns into the isotropic gapped
one,
\begin{equation}\label{EK}
\tilde{\varepsilon}=\pm\sqrt{(\varepsilon_g/2)^2+(\hbar vk)^2},
\end{equation}
where
$\varepsilon_g=\sqrt{(2veE_0/\omega)^2+(\hbar\omega)^2}-\hbar\omega$
is the field-induced gap \cite{Kibis_10}. Combining the theory of
gapped graphene subjected to a magnetic field \cite{Koshino_10}
and the theory of graphene layer dressed by a circularly polarized
electromagnetic wave \cite{Kibis_10}, we easily arrive from the
Hamiltonian (\ref{H0}) at the energy spectrum of Landau levels in
the $K$ point of Brillouin zone,
\begin{equation}\label{EM}
\tilde{\varepsilon}_n=\pm\sqrt{(\hbar\omega_B)^2(n+1/2\mp1/2)+\varepsilon_g^2},
\end{equation}
where $n=0,1,2,...$ is the number of Landau level, and the upper
and lower signs correspond to the conductivity and valence band,
respectively. As to the Landau levels at the $K^\prime$ point,
their structure can be also described by Eq.~(\ref{EM}), where the
signs ``$\pm$'' under the square root should be replaced with the
opposite signs, ``$\mp$''. It should be noted that the energy
spectrum (\ref{EM}) is correct for weak magnetic fields satisfying
the condition $\hbar\omega_B\ll\varepsilon_g$ (for the opposite
case of strong magnetic fields see Ref.~\cite{Lopez_15}). In
contrast to Eq.~(\ref{En}), the energy spectrum (\ref{EM}) does
not contain the Bessel function, $J_0(\alpha)$. Mathematically,
this means that the dependence of the magnetoelectronic properties
of graphene on a dressing field is more pronounced in the case of
linear polarization. Thus, a linearly polarized dressing field is
preferable from experimental viewpoint.

As to experimental observability of the discussed effects, it
should be noted that the employed parameters of the dressing field
(the photon energy of meV scale and the field intensity of
W/cm$^2$) can be easily realized experimentally with using such
conventional sources of THz radiation as quantum cascade lasers
and free electron lasers (see, e.g.,
Refs.~\cite{Williams_07,Carr_02}).

Summarizing the aforesaid, we can conclude that a strong
high-frequency electromagnetic field (dressing field)
substantially modifies magnetoelectronic properties of graphene in
contrast to the case of conventional conductors with the parabolic
dispersion of electrons. Particularly, such resonant effects as
optical absorption and Shubnikov-de Haas oscillations are very
sensitive to the field. Therefore, a dressing field can be
considered as a perspective tool to manipulate the
magnetoelectronic properties of graphene. Since graphene serves as
a basis for nanoelectronic devices, the developed theory opens a
new way to control their characteristics.

\section{Acknowledgments} The work was partially supported by FP7 IRSES project QOCaN, FP7
ITN project NOTEDEV, Rannis project BOFEHYSS, Singaporean Ministry
of Education under AcRF Tier 2 grant MOE2015-T2-1-055. OVK thanks
RFBR project 14-02-00033 and Russian Ministry of Education and
Science for the support. IAS thanks the Russian Target Federal
Program ``Research and Development in Priority Areas of
Development of the Russian Scientific and Technological Complex
for 2014-2020'' of the Ministry of Education and Science of Russia
(project 14.587.21.0020) for the support.

\appendix
\section{Borders of applicability of the theory}

In order to simplify comparison of the developed theory and
experiments, we summarize the conditions of applicability of the
theory:

(i) The dressing field is assumed to be off-resonant. Therefore,
the frequency of the dressing field, $\omega$, should be far from
resonant frequencies corresponding to electron transitions between
different Landau levels;

(ii) To exclude the scattering-induced intra-level absorption of
the wave by electrons within a broadened Landau level, we have to
assume the wave frequency, $\omega$, to be high enough in order to
satisfy the inequality $\omega\tau\gg1$.  It is well-known that
the nonresonant (collisional) absorption of wave energy by
conduction electrons is negligibly small under this condition.
Therefore, a dressing field which satisfies both the nonresonant
condition (i) and the high-frequency condition (ii) can be treated
as a purely dressing (non-absorbable) field;

(iii) The effective Hamiltonian (\ref{H1}) is derived by reducing
the infinite system of quantum dynamics (\ref{a1}) to the sole
equation (\ref{a2}). This reducing is correct under the condition
$|J_n(\alpha)/J_0(\alpha)|\ll1$, where $J_n(\alpha)$ is the Bessel
function of the first kind, $\alpha={2veE_0}/{\hbar\omega^2}$ is
the dimensionless parameter describing the interaction between an
electron in graphene and a dressing field, $v$ is the electron
velocity in graphene, $E_0$ is the amplitude of the dressing
field, $\omega$ is the frequency of the dressing field, and
$n=\pm1,\pm2,\pm3...$. Thus, the dressing field amplitude, $E_0$,
and the field frequency, $\omega$, should be chosen to keep the
Bessel function, $J_0(\alpha)$, far from zero;

(iv) The calculation of the conductivity
(\ref{Kubo1})--(\ref{Kubo}) within the approach \cite{Ando_98} was
performed under the condition $\Gamma/\varepsilon_F\ll1$, where
$\Gamma$ is the scatterer-induced broadening of Landau levels, and
$\varepsilon_F$ is the Fermi energy. Therefore, the developed
theory of transport effects is adequate if the scattering is weak
enough.


\begin{thebibliography}{99}

\bibitem{Novoselov_04}
K. S.~Novoselov, A. K.~Geim, S. V.~Morozov, D.~Jiang, Y.~Zhang, S.
V.~Dubonos, I. V.~Grigorieva, A. A.~Firsov, Science {\bf 306}, 666
(2004).

\bibitem{CastroNeto_2009}
A. H. Castro Neto, F. Guinea, N. M. R. Peres, K. S. Novoselov, A.
K. Geim,  {Rev. Mod. Phys.} {\bf 81}, 109 (2009).

\bibitem{DasSarma_2011}
S. Das Sarma, S. Adam, E. H. Hwang, E. Rossi, {Rev. Mod. Phys.}
{\bf 83}, 407 (2011).

\bibitem{Goerbig}
M. O. Goerbig, Rev. Mod. Phys. {\bf 83}, 1193 (2011).

\bibitem{Yang}
C. H. Yang, F. M. Peeters, W. Xu, Phys. Rev. B {\bf 82}, 075401
(2010).

\bibitem{Jiang}
A. Luican, G. Li, E. Y. Andrei, Phys. Rev. B {\bf 83}, 041405
(2011).

\bibitem{Zhang}
F. Libisch, S. Rotter, J. Guttinger, C. Stampfer, J. Burgdorfer,
Phys. Rev. B {\bf 81}, 245411 (2010).

\bibitem{Sadowski}
N. Gu, M. Rudner, A. Young, P. Kim, L. Levitov, Phys. Rev. Lett.
{\bf 106} , 066601 (2011).

\bibitem{Yao}
X. Yao, A. Belyanin, Phys. Rev. Lett. {\bf 108}, 255503 (2012).

\bibitem{Booshehri}
L. G. Booshehri, C. H. Mielke, D. G. Rickel, S. A. Crooker, Q.
Zhang, L. Ren, E. H. Haroz, A. Rustagi, C. J. Stanton, Z. Jin, Z.
Sun, Z. Yan, J. M. Tour, J. Kono, Phys. Rev. B {\bf 85}, 205407
(2012).

\bibitem{Crassee}
I. Crassee, J. Levallois, A. L. Walter, M. Ostler, A. Bostwick, E.
Rotenberg, T. Seyller, D. van der Marel, A. B. Kuzmenko, Nature
Phys. {\bf 7}, 48 (2011).

\bibitem{Crasseebis}
I. Crassee, M. Orlita, M. Potemski, A. L. Walter, M. Ostler, Th.
Seyller, I. Gaponenko, J. Chen, A. B. Kuzmenko,  Nano Lett. {\bf
12}, 2470 (2012).

\bibitem{Grujic}
M. Gruji\`c, M. Zarenia, A. Chaves, M. Tadic, G. A. Farias, F. M.
Peeters, Phys. Rev. B {\bf 84}, 205441 (2011).

\bibitem{Shimano}
R. Shimano, G. Yumoto,  J. Y. Yoo,  R. Matsunaga,   S. Tanabe,  H.
Hibino,  T. Morimoto , H. Aoki, Nature Commun. {\bf 4}, 1841
(2013).

\bibitem{Yao_13}
X. Yao, A. Belyanin, J. Phys: Condens. Matter {\bf 25} 054203
(2013).

\bibitem{Ando_98} N. H. Shon, T. Ando, J. Phys. Soc. Japan {\bf
67}, 2421 (1998).

\bibitem{Guinea}
F. Guinea, M. I. Katsnelson, A. K. Geim, Nature Phys. {\bf 6}, 30
(2010) .

\bibitem{Das_Sarma}
S. Das Sarma, S. Adam, E. H. Hwang, Enrico Rossi, Rev. Mod. Phys.
{\bf 83}, 407 (2011).

\bibitem{Tan}
Z. Tan, C. Tan, L. Ma, G. T. Liu, L. Lu, C. L. Yang, Phys. Rev. B
{\bf 84}, 115429 (2011).

\bibitem{Qiao}
Z. Qiao, S. A. Yang, W. Feng, W. K. Tse, J. Ding, Y. Yao, J. Wang,
Q. Niu, Phys. Rev. B {\bf 82}, 161414 (2010).

\bibitem{Waldmann}
J. Jobst, D. Waldmann, F. Speck, R. Hirner, D. K. Maude, T.
Seyller, H. B. Weber, Phys. Rev. B {\bf 81}, 195434 (2010).

\bibitem{Waldmannbis}
D. Waldmann, J. Jobst, F. Speck, T. Seyller, M. Krieger, H. B.
Weber, Nature Mater. {\bf 10}, 357 (2011).

\bibitem{Dean}
C. R. Dean, A. F. Young, P. Cadden-Zimansky, L. Wang, H. Ren, K.
Watanabe, T. Taniguchi, P. Kim, J. Hone, K. L. Shepard, Nature
Phys. {\bf 7} , 693 (2011).

\bibitem{Cohen-Tannoudji_b98}
C. Cohen-Tannoudji, J. Dupont-Roc, G. Grynberg,
\textit{Atom-Photon Interactions: Basic Processes and
Applications}  (Wiley, Chichester, 1998).

\bibitem{Scully_b01}
M. O. Scully, M. S. Zubairy, \textit{Quantum Optics} (Cambridge
University Press, Cambridge, 2001).

\bibitem{Autler_55}
S. H. Autler, C. H. Townes, Phys. Rev. {\bf 100}, 703 (1955).

\bibitem{Elesin_69}
S. P. Goreslavskii, V. F. Elesin, JETP Lett. {\bf 10}, 316 (1969).

\bibitem{Vu_04}
Q. T. Vu, H. Haug, O. D. M\"ucke, T. Tritschler, M. Wegener, G.
Khitrova, H. M. Gibbs, Phys. Rev. Lett. \textbf{92}, 217403
(2004).

\bibitem{Vu_05}
Q. T. Vu, H. Haug, Phys. Rev. B {\bf 71}, 035305 (2005).

\bibitem{Mysyrovich_86}
A. Myzyrowicz, D. Hulin, A. Antonetti, A. Migus, W. T. Masselink,
H. Morko\c{c}, Phys. Rev. Lett. {\bf 56}, 2748 (1986).

\bibitem{Wagner_10}
M. Wagner, H. Schneider, D. Stehr, S. Winnerl, A. M. Andrews, S.
Schartner, G. Strasser, M. Helm, Phys. Rev. Lett. {\bf 105},
167401 (2010).

\bibitem{Kibis_12}
O. V. Kibis, Phys. Rev. B {\bf 86}, 155108 (2012).

\bibitem{Teich_13}
M. Teich, M. Wagner, H. Schneider, M. Helm, New J. Phys. {\bf 15},
065007 (2013).

\bibitem{Kibis_14}
O. V. Kibis, Europhys. Lett. {\bf 107}, 57003 (2014).

\bibitem{Morina_15}
S. Morina, O. V. Kibis, A. A. Pervishko, I. A. Shelykh, Phys. Rev.
B {\bf 91}, 155312 (2015).

\bibitem{Pervishko_15}
A. A. Pervishko, O. V. Kibis, S. Morina, I. A. Shelykh, Phys. Rev.
B {\bf 92}, 205403 (2015).

\bibitem{Kibis_11}
O. V. Kibis, Phys. Rev. Lett. {\bf 107}, 106802 (2011).

\bibitem{Kibis_14_1} O. V. Kibis, O. Kyriienko, I. A. Shelykh, Phys. Rev. B {\bf 87},
245437 (2013).

\bibitem{Joibari_14}
F. K. Joibari, Y. M. Blanter, G. E. W. Bauer, Phys. Rev. B {\bf
90}, 155301 (2014).

\bibitem{Kyriienko_15}
G. Yu. Kryuchkyan, O. Kyriienko, I. A. Shelykh, J. Phys. B {\bf
48}, 025401 (2015).

\bibitem{Lopez_08}
F. J. L\'{o}pez-Rodr\'{i}guez, G. G. Naumis, Phys. Rev. B {\bf
78}, 201406(R) (2008).

\bibitem{Oka_09}
T.~Oka and H.~Aoki, \prb {\bf 79}, 081406(R) (2009).

\bibitem{Kitagawa_11}
T. Kitagawa, T. Oka, A. Brataas, L. Fu, E. Demler, Phys. Rev. B
{\bf 84}, 235108 (2011).

\bibitem{Kibis_10}
O. V. Kibis, Phys. Rev. B {\bf 81}, 165433 (2010).

\bibitem{Kibis_11_1}
O. V. Kibis, O. Kyriienko, I. A. Shelykh, Phys. Rev. B {\bf 84},
195413 (2011).

\bibitem{Usaj_14}
G. Usaj, P. M. Perez-Piskunow, L. E. F. Foa Torres, C. A.
Balseiro, Phys. Rev. B {\bf 90}, 115423 (2014).

\bibitem{Glazov_14}
M. M. Glazov, S. D. Ganichev, Phys. Rep. {\bf 535}, 101 (2014).

\bibitem{Lopez_15}
A. L\'{o}pez, A. Di Teodoro, J. Schliemann, B. Berche, B. Santos,
Phys. Rev. B {\bf 92}, 235411 (2015).

\bibitem{Zeldovich_67}
Ya. B. Zel'dovich, Sov. Phys. JETP {\bf 24}, 1006 (1967).

\bibitem{Grifoni_98}
M. Grifoni, P. H\"anggi, Phys. Rep. {\bf 304}, 229 (1998).

\bibitem{Platero_04}
G. Platero, R. Aguado, Phys. Rep. {\bf 395}, 1 (2004).

\bibitem{Kristinsson_16}
K. Kristinsson, O. V. Kibis, S. Morina, I. A. Shelykh, Sci. Rep.
{\bf 6}, 20082 (2016).

\bibitem{Inoshita_00}
T. Inoshita, Phys. Rev. B {\bf 61}, 15610 (2000).

\bibitem{Koshino_10}
M. Koshino, T. Ando, Phys. Rev. B {\bf 81}, 195431 (2010).

\bibitem{Williams_07}
B. S. Williams, Nature Photonics {\bf 1}, 517 (2007).

\bibitem{Carr_02}
G. L. Carr, M. C. Martin, W. R. McKinney, K. Jordan, G. R. Neil,
G. P. Williams, Nature {\bf 420}, 153 (2002).

\end{thebibliography}
\end{document}